\documentclass[apj]{emulateapj}
\usepackage{apjfonts,hyperref,breakurl}
\submitted{Received 2008 October 9; accepted 2009 August 4}
\journalinfo{Accepted version 2009 August 4}

\newcommand{\kms}{km~s$^{-1}$}
\newcommand{\ha}{H$\alpha$}
\newcommand{\dt}{$\delta$}
\newcommand{\sm}{$\sim$}
\newcommand{\dg}{$^{\circ}$}

\newcommand{\ovsa}{Owens Valley Solar Array}
\newcommand{\Hsi}{\textit{Reuven Ramaty High Energy Solar Spectroscopic Imager}}
\newcommand{\hsi}{\textit{RHESSI}}
\newcommand{\Trace}{\textit{Transition Region and Coronal Explorer}}
\newcommand{\trace}{\textit{TRACE}}
\newcommand{\Soho}{\textit{Solar and Heliospheric Observatory}}
\newcommand{\soho}{\textit{SOHO}}
\newcommand{\goes}{\textit{GOES}}

\newcommand{\wind}{\textit{WIND}}

\newcommand{\lasco}{Large Angle and Spectrometric Coronagraph}
\newcommand{\mdi}{Michelson Doppler Imager}
\newcommand{\Spirit}{\textit{SPectroheliographIc X-Ray Imaging Telescope}}
\newcommand{\spirit}{\textit{SPIRIT}}

\begin{document}
\title{SUCCESSIVE SOLAR FLARES AND CORONAL MASS EJECTIONS ON 2005 SEPTEMBER 13 FROM NOAA AR 10808}
\author{CHANG LIU\altaffilmark{1}, JEONGWOO LEE\altaffilmark{2}, MARIAN KARLICK{\'Y}\altaffilmark{3}, DEBI PRASAD CHOUDHARY\altaffilmark{4}, NA DENG\altaffilmark{4},\\AND HAIMIN WANG\altaffilmark{1}}
\affil{1. Space Weather Research Laboratory, Center for Solar-Terrestrial Research, New Jersey Institute of Technology, University Heights,\\Newark, NJ 07102, USA; chang.liu@njit.edu, haimin@flare.njit.edu}
\affil{2. Physics Department, New Jersey Institute of Technology, University Heights, Newark, NJ 07102-1982, USA; leej@njit.edu}
\affil{3. Astronomical Institute of the Academy of Sciences of the Czech Republic, 25165 Ond{\v r}ejov, Czech Republic; karlicky@asu.cas.cz}
\affil{4. Department of Physics and Astronomy, California State University, Northridge, CA 91330-8268, USA;\\debiprasad.choudhary@csun.edu, na.deng@csun.edu}

\shorttitle{SUCCESSIVE SOLAR FLARES AND CORONAL MASS EJECTIONS}
\shortauthors{LIU ET AL.}

\begin{abstract}
We present a multiwavelength study of the 2005 September 13 eruption
from NOAA 10808 that produced total four flares and two fast coronal mass ejections (CMEs) within \sm1.5 hours. Our primary attention is paid to the fact that these eruptions
occurred in close succession in time, and that all of them were
located along an S-shaped magnetic polarity inversion line (PIL) of
the active region. In our analysis, (1) the disturbance created 
by the first flare propagated southward along the PIL to cause a major
filament eruption that led to the first CME and the associated second flare underneath. (2) The first CME partially removed the overlying magnetic fields over the northern \dt\ spot
to allow the third flare and the second CME. (3) The ribbon separation
during the fourth flare would indicate reclosing of the overlying field lines
opened by the second CME. It is thus concluded that this series of flares 
and CMEs are interrelated to each other via magnetic reconnections between
the expanding magnetic structure and the nearby magnetic fields. These results complement previous works made on this event with the suggested causal relationship among the successive eruptions.

\end{abstract}

\section{INTRODUCTION}
Although solar flares are known as the process of sudden energy
release in a restricted volume of the solar atmosphere, some of them
are not isolated events. Statistical studies on the flare occurrence
have suggested a sympathetic flaring activity, in which flares occur
almost simultaneously in different active regions
\citep{richardson51,fritzova76,pearce90,moon02,wheatland06a}.
Several case studies followed to show how different sites of
sympathetic flares are connected to each other, for instance, by way of X-ray ejecta \citep{gopalswamy99},
magnetic reconnections with neighboring fields \citep{bagala00},
heat conduction along large-scale loops \citep{zhang00}, and/or
sweeping close-loop surge \citep{wang01}. An MHD simulation demonstrated that the sympathetic flares can also be
triggered by shocks or high energy particle beams \citep{odstrcil97}. The idea of sympathetic flares has been
integrated into a model for solar flare statistics
\citep{wheatland06b}. Since coronal mass ejections (CMEs) are
closely related to flares, it is natural to expect that there also
exist sympathetic CMEs, i.e., a pair of consecutive CMEs
originating from different active regions and physically connected
with each other. However, as implied by statistical results of
waiting-time distribution and the angular-difference distribution,
the number of sympathetic CMEs is much smaller than that of
independent CMEs \citep{moon03}. Only a few cases of sympathetic
CMEs have been reported thus far \citep{simnett97,cheng05,jiang08},
and their possible driving mechanism is still an area of ongoing
research.

There is another class of activity of multiple flares/CMEs.
\citet{goff07} studied a series of three flares (two M and one C in
\goes\ class) and one CME occurred within \sm1 hour in NOAA AR 10540
on 2004 January 20. They suggested that the CME is related to the
first two flares via tether-cutting reconnection \citep{moore01},
while the third flare in a close-by location is caused by
interaction between the expanding CME and neighboring magnetic
fields. \citet{gary04} examined a spiral flux tube eruption from
NOAA AR 10030 on 2002 July 15 in the context of the breakout model
\citep{antiochos99}. Interestingly, after the first CME and an
associated X3.1 flare, a second M4.1 flare and another CME were
subsequently released from a nearby stressed fields within \sm30
minutes, which was interpreted as due to decrease of the overlying
fields as a result of the first magnetic breakout. This kind of multiple
eruption activity differs from the sympathetic flares/CMEs, because
the term \textit{sympathetic} historically alludes to the events
involving different active regions. Instead we will call them
\textit{successive} flares/CMEs in the sense that multiple eruptions occur in one active region within a short time period. Understanding of the successive
flares/CMEs would certainly be important for solar flare physics and
for space weather, since it will shed light on the physics of
initiation and mechanisms of energy transport of solar eruptions.

\begin{deluxetable}{ccccc}
\tablewidth{0pt}
\tablecaption{DATA FOR 13-SEP-2005 NOAA X1.5 FLARE-CME EVENT\label{data}}
\tablehead{\colhead{} & \colhead{Data Type/} & \colhead{Coverage} & \colhead{Cadence} & \colhead{Spatial}\\
\colhead{Instrument} & \colhead{Band} & \colhead{(UT)} & \colhead{(seconds)} & \colhead{Resolution (\arcsec)}}
\startdata
OSPAN      & \ha, $\pm$0.4~\AA\ & Full      &   \sm60       & 2.1     \\
BBSO	   & \ha~$-$~0.6~\AA\  & Full       &    30         & 2.1     \\ 
MDI        & magnetogram       & Full       &    60         & 4       \\
\hsi\      & X-rays            & \sm19:54-- &    4          & up to 2.3 \\
\trace\    & 195~\AA\          & \sm19:24-- &  \sm15        & 2       \\
\spirit\   & 175~\AA\          & Partial    & few images    & 10.2    \\
LASCO      & white-light       & Full       & \sm12 mins    & 24/112  \\
OVSA       & 1.2--18~GHz       & Full       &    8.1        & 4.9     \\
GBSRBS     & 18--1057~MHz      & Full       &     1         & -       \\
WAVES      & 4~kHz--14~MHz      & Full      &    16         & -
\enddata
\tablecomments{OSPAN: USAF/Optical Solar Patrol Network \citep[formerly known as ISOON;][]{neidig98}; BBSO: Big Bear Solar Observatory \citep{wang98}; MDI: \mdi\ \citep{scherrer95} on the \Soho\ (\soho); \hsi: \Hsi\ \citep{lin02}; \trace: \Trace\ \citep{handy99}; \spirit: \Spirit\ \citep{slemzin05}; LASCO: \lasco\ \citep{brueckner95}; OVSA: \ovsa\ \citep{gary90}; GBSRBS: Green Bank Solar Radio Burst Spectrometer \citep{bastian05}; WAVES: \wind\ Radio and Plasma Wave Experiment \citep{bougeret95}.}
\end{deluxetable}

In this paper, we study the multiple eruptions of flares and CMEs
during \sm19--21~UT on 2005 September 13 from the NOAA AR 10808,
which appears to be an excellent candidate for the successive
flares/CMEs. The plan of this paper is as follows: in \S\S~\ref{previous} and \ref{overview}, we first summarize related previous studies, and overview the active region and the event timing. In \S\S~\ref{flares} and \ref{cme} the multiwavelength
observations of the whole event are described. We speculate on
plausible magnetic reconnection scenarios in \S~\ref{idea}, and
summarize and discuss our major findings in \S~\ref{summary}.

\section{Previous Studies on the Event} \label{previous}

The source active region, NOAA 10808, had produced extraordinary
eruption activities during the solar cycle 23, and many eruptions
from this active region have already received attention from many
authors. For the purpose of this study, it is worthwhile to review
the published results related to this event.

First, \citet{chifor07} carried out a study of X-ray precursors and
filament eruptions in eight events including the present one. They
found traveling hard X-ray (HXR) brightenings in preflare times and
regarded them as a precursor of filament eruption and the main
energy release. Second, \citet{nagashima07} studied a series of
small flares occurred in adjacent regions over two days before the
eruption. They found the slow ascending motion of a major filament
and concluded that the small flares led to the eruption of the
filament. Third, \citet{wang07a} measured the relative timing of
flare ribbons and the filament, and suggested that the filament
eruption occurred as the flux loop system is destabilized by the
initial flare in a nearby location. In addition, \citet{wang06b}
carried out a large-scale modeling for five CMEs from this
active region. \citet{li07} presented the magnetic configuration
of the active region at the time of another flare that occurred \sm4 hours later.

The present study is motivated by the fact that the whole event is
more complex than described in the above studies. Within the
duration of \sm1.5 hours, there occurred consecutively four flares
and two CMEs, and it is necessary to further study possible relationships among them. In particular, we will show an
additional filament eruption at the beginning of the event that was overlooked in the previous studies. This affects the
interpretations regarding the initiation of the whole eruption. We
will also present additional flares that were not studied in the
above papers. We intend to revisit the initiation scenario of each eruption
and possible interrelationships among them, using nearly all
available ground-based and space-borne multiwavelength data from ten
observational instruments (Table~\ref{data}).

\begin{figure}
\epsscale{1.15}
\plotone{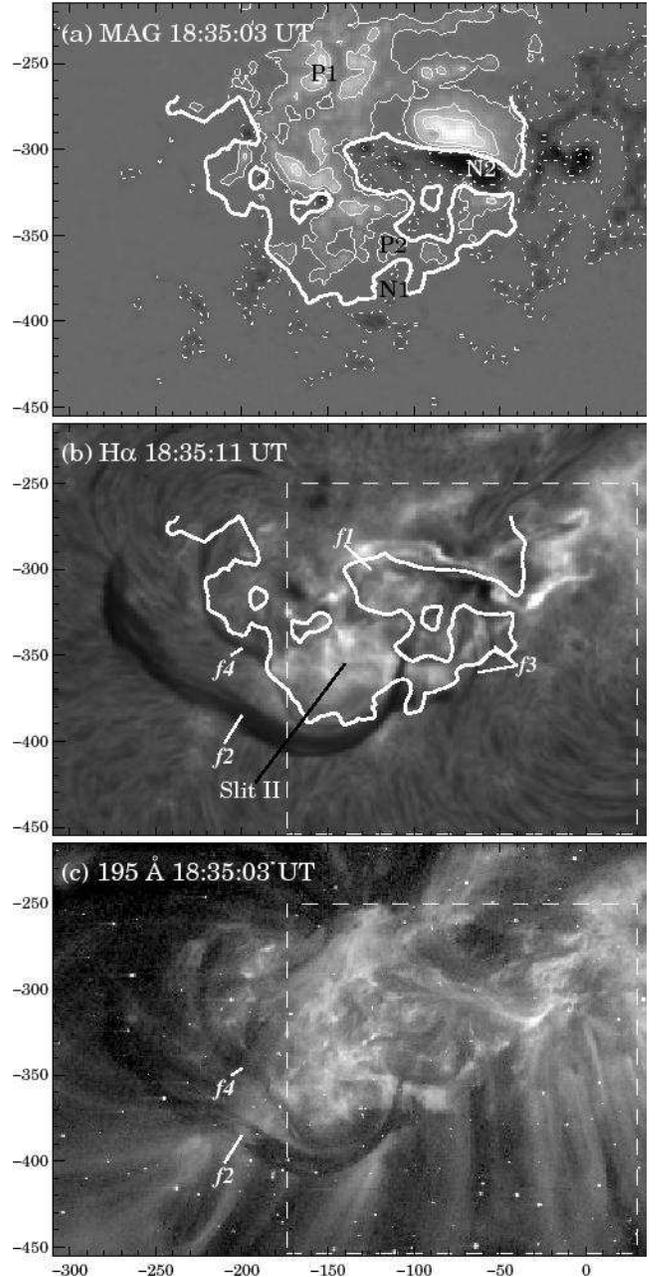}
\caption{{\it Top}: An MDI magnetogram showing the principal \dt\ spot, surrounding magnetic islands, and an overall S-shaped PIL (thick white line) of NOAA AR 10808 on 2005 September 13. P1, P2, N1, and N2 denote magnetic regions involved with the CME I (see \S~\ref{first}). {\it Middle}: An OSPAN preflare \ha\ image showing the erupted filaments $f1$, $f2$, and $f3$ and an undisturbed filament $f4$. The black line indicates the slit position of the time slice shown in Fig.~\ref{slit}$b$. {\it Bottom}: A \trace\ 195~\AA\ image showing the preflare coronal conditions. Axis scales are in arcsec. West is to the right, and solar north is up. The dashed box indicates the field of view of \ha\ images in Fig.~\ref{fig2}$a$. All the images as well as those in Figs.~\ref{fig2} are aligned respect to 20:01~UT.} \label{fig0}
\end{figure}

\begin{figure}
\epsscale{1.15}
\plotone{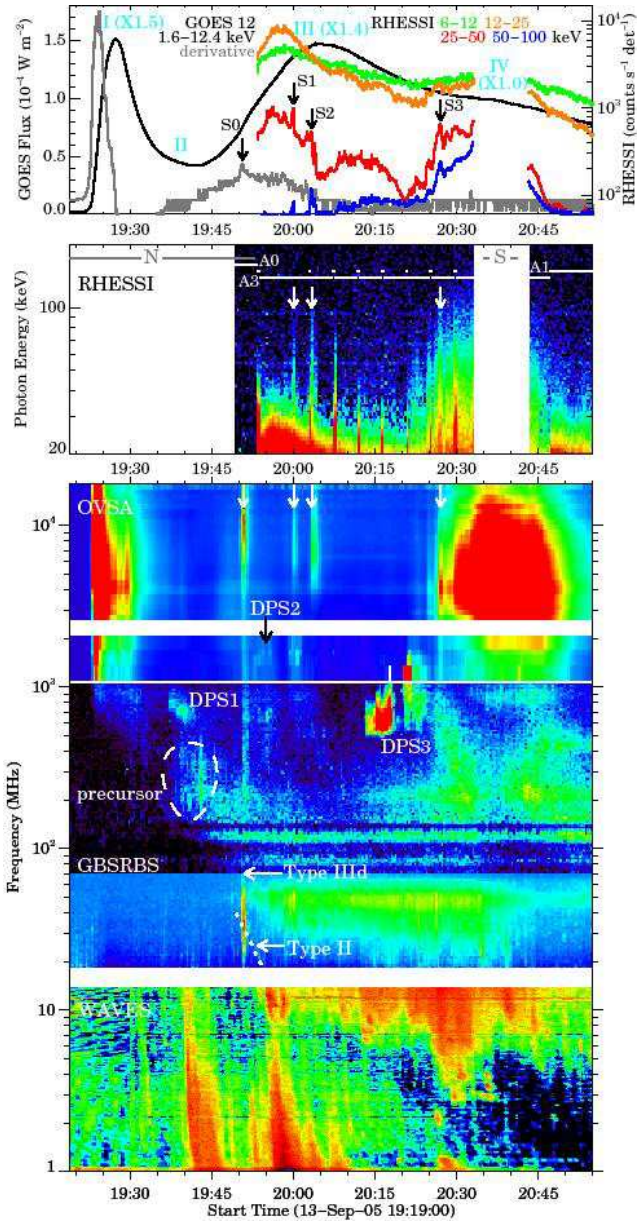}
\caption{\textit{Top panel}: Time evolution of the \goes\ SXR flux and its time derivative (smoothed with a width of 12~s) overplotted with \hsi\ corrected count rates. Progression of flares I--IV are illustrated. $S0$--$S3$ denote several high energy bursts characterizing the flares III and IV. \textit{Second panel}: \hsi\ dynamic spectrum in 20--200~keV. N and S denote the time period of \hsi\ night and South Atlantic anomaly, respectively. Some periodic spikes during \sm19:50--20:30~UT are due to the noise when \hsi\ attenuator switched between A1 and A3 status. \textit{Third to bottom panels}: Composite radio dynamic spectrum observed by OVSA, GBSRBS, and WAVES. The dotted line marks the type II radio burst, which is clearly visible in an enlarged spectrum. The type II precursor, type IIId burst, and several DPSs are also indicated. See text in \S\S~\ref{overview} and \ref{radio} for details. \label{fig1}}
\end{figure}

\section{OVERVIEW OF THE EVENT} \label{overview}
As an overview, we show, in Figure~\ref{fig0}, the magnetogram, \ha,
and extreme-UV (EUV) images of the active region, where we identify the filaments
involved with the eruption. In Figure~\ref{fig1}, we then present light
curves and dynamic spectra of X-rays and radio wavelengths where we
identify the individual flare peaks.

The top panel of Figure~\ref{fig0} shows an MDI magnetogram of the source
active region NOAA 10808. It contains a complex bipolar sunspot
group with one main $\delta$ spot, and appears in the
$\beta\gamma\delta$ configuration. The overall magnetic polarity
inversion line (PIL) is S-shaped, which is located close to disk center
(S11\dg, E6\dg) at the time of this event. Its long-term evolution
shows that the principal \dt\ spot underwent a fast counterclockwise
rotation, which implies strongly sheared and twisted magnetic fields
there \citep{li07}. In Figure~\ref{fig0} ({\it middle}) an OSPAN
preflare \ha\ image shows well the filaments that are closely
related to the present study. The filament $f1$ lies in the
northeast corner of the PIL, $f2$ and $f4$ around the southeast
portion of the PIL, and $f3$ just south of the \dt\ spot. Most of
these filaments are also visible in a preflare 195~\AA\ coronal
image (Fig.~\ref{fig0}, {\it bottom}). A gradual ascending motion of
$f2$ with a series of small flares occurred nearby over two days
before the eruption was observed by \citet{nagashima07}.

In Figure~\ref{fig1} we show the light curves and dynamic spectra of
X-rays and radio wavelengths, which were largely missing from
previous studies. According to the \goes\ soft X-ray (SXR) flux, the
event started at 19:19~UT, peaked at 19:27 and 20:05 (X1.5 and X1.4,
hereafter referred as the first [I] and the third [III] flare), and
ended at 20:57~UT. Note that NOAA listed the event as a single X1.5
flare. We will see later that there occurred a second (II) flare
around 19:35~UT but it can not be clearly identified in SXR flux.
Moreover, it is obvious in this figure that an additional emission
episode occurred after \sm20:21~UT (X1.0, hereafter referred as the
fourth [IV] flare). This kind of consecutive multiple SXR peaks in a
short time period with corresponding nonthermal emissions is a
typical characteristic of successive flare/CME event \citep[e.g.,][]{gary04}.

We also indicated, with arrows,
the four nonthermal bursts, $S0$--$S3$, seen in 25--100~keV HXRs and/or
microwaves at \sm19:51, 20:00, 20:04, and 20:27~UT, respectively. The
peak $S0$ is inferred based on the time derivative of SXR flux
\citep{neupert68}, as HXR data are not available at that time. $S1$
peaks simultaneously in both HXRs and microwave at
\sm20:00:04~UT. Although $S2$ is partially affected by the noise
when the \hsi\ attenuator switched between A1 and A3 status during
20:02:48--20:03:12~UT, the 25--100~keV energy range has a peak at
20:03:22~UT and could be associated with a delayed peak in microwave
at 20:03:25~UT. $S3$ peaks in HXR and microwaves at 20:27:04 and
20:27:10, respectively. Therefore, we believe that each of these
four bursts have corresponding peaks in both HXRs and microwaves. In
particular, $S0$ is at the beginning of the impulsive phase of the
flare III and is also a broadband type IIId burst
\citep{poquerusse94} spanning from 1~MHz to 18~GHz, which is most
probably associated with an ordinary type III burst starting at
\sm19:55~UT, as discussed in \citet{klassen03}. Other radio
signatures associated with magnetic eruption will be discussed in
\S~\ref{radio}.

\begin{figure*}
\epsscale{1.02}
\plotone{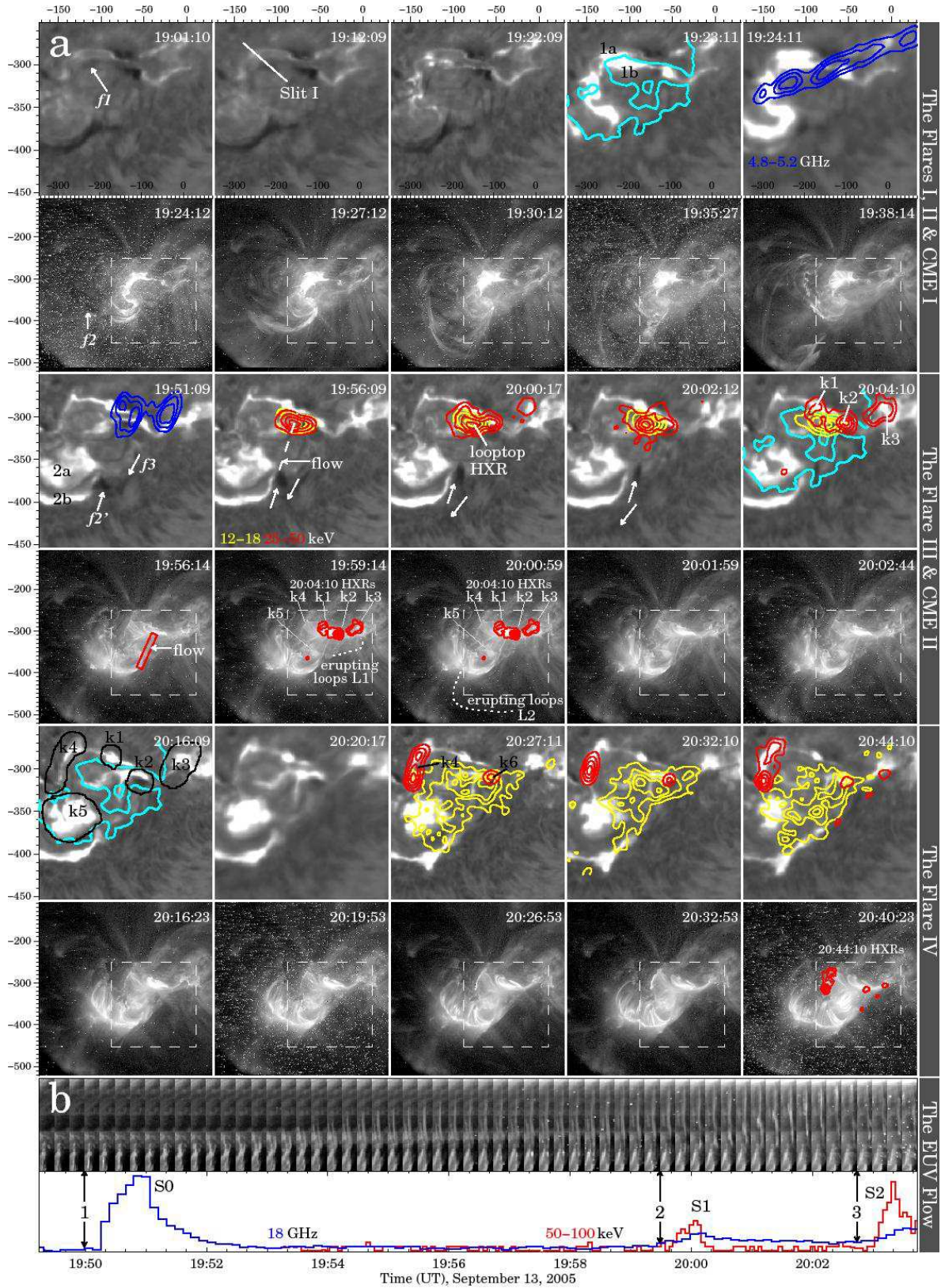}
\caption{\textit{(a)} Time sequence of OSPAN \ha~$-$~0.4~\AA\ and \trace\ 195~\AA\ images showing the evolution of the event, with cleaned \hsi\ X-ray and OVSA microwave sources superposed. Each \hsi\ image was reconstructed using detectors 3--8 (9.8\arcsec\ FWHM resolution). The integration time of both X-ray and microwave images is 1 minute centering on the corresponding \ha\ times. Contour levels are 30\%, 50\%, 70\%, and 90\% of the maximum flux for \hsi, and 40\%, 60\%, and 80\% for OVSA. The dashed box in the EUV images represents the field of view of \ha\ images. The white line in the image of 19:12:09~UT indicates the slit position of the time slice shown in Fig.~\ref{slit}$a$. The turquoise line denotes the PIL. \textit{(b)} Time sequence combined \trace\ 195~\AA\ images using a small area (15\arcsec~$\times$~90\arcsec) around the EUV flow (illustrated in the frame 19:56:14~UT), which is rotated 24\dg\ counterclockwise. Arrows 1--3 mark the initiation of different flow episodes, which seem to co-temporal with flare nonthermal emissions in HXR and microwave. \label{fig2}}
\end{figure*}

\section{THE SUCCESSIVE FLARES} \label{flares}
We show the flare evolution observed at various frequencies in Figure~\ref{fig2}$a$, and describe the results in this section. More dynamic details can be seen in the accompanying mpeg movies\footnote{\url{http://solar.njit.edu/~cliu/050913}}.

\begin{figure}
\epsscale{1.15}
\plotone{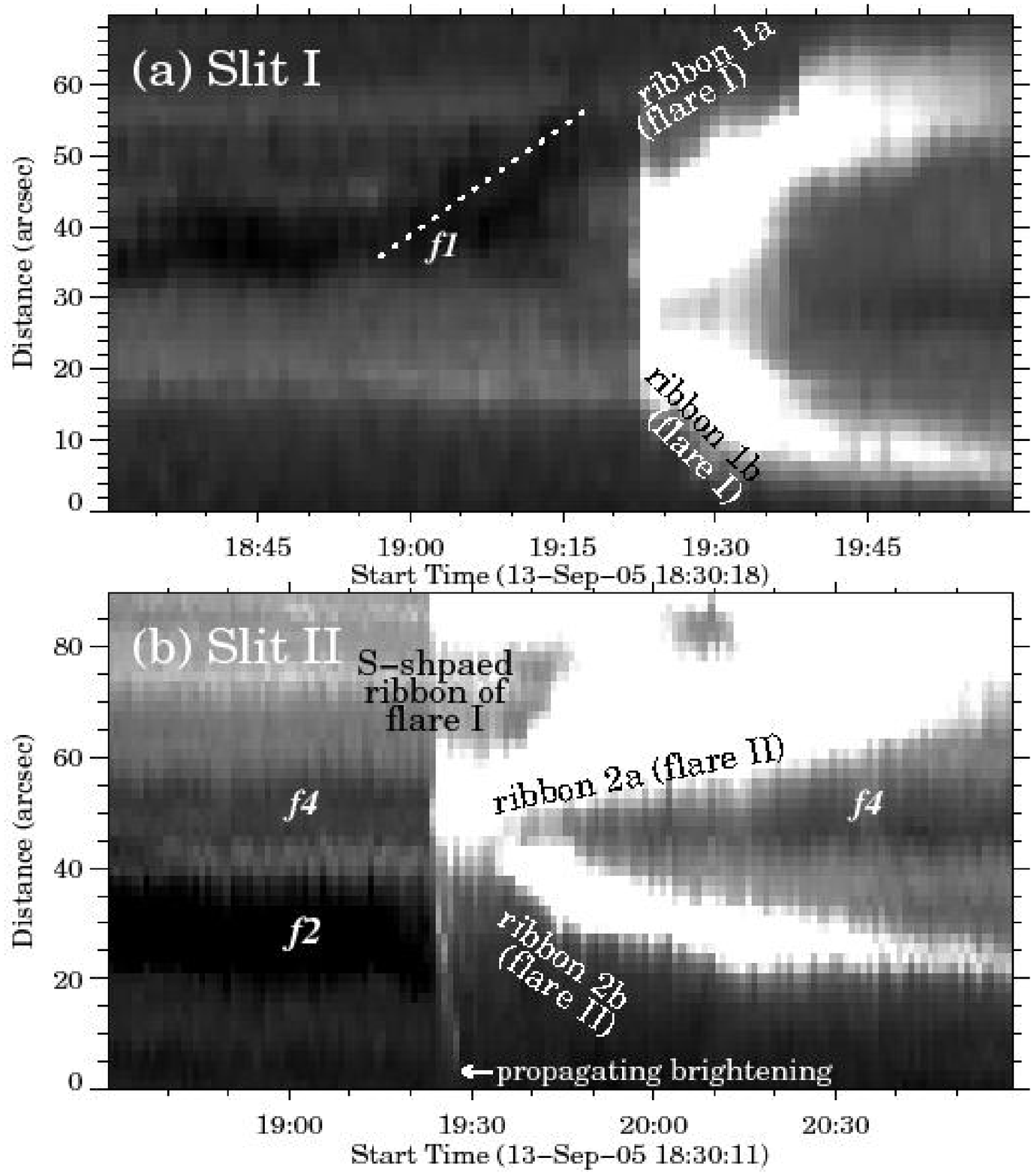}
\caption{Time slices for the slit I (see the 19:12:09 frame in Fig.~\ref{fig2}$a$) using \ha~$-$~0.4~\AA\ images and the slit II (see Fig.~\ref{fig0}$b$) using \ha\ center images. The distance is measured from the northern and southern ends of the slit I and II, respectively. The orientation of the slits were chosen in order to show clearly the dynamics of the filaments $f1$, $f2$, and $f4$. Flare ribbons $1a$/$1b$ and $2a$/$2b$ are indicated in the frames 19:23:11 and 19:51:09~UT in Fig.~\ref{fig2}$a$, respectively. \label{slit}}
\end{figure}

\subsection{The Flares I and II (\sm19:19--19:45~UT)} \label{initial}
\begin{figure*}
\epsscale{1.15}
\plotone{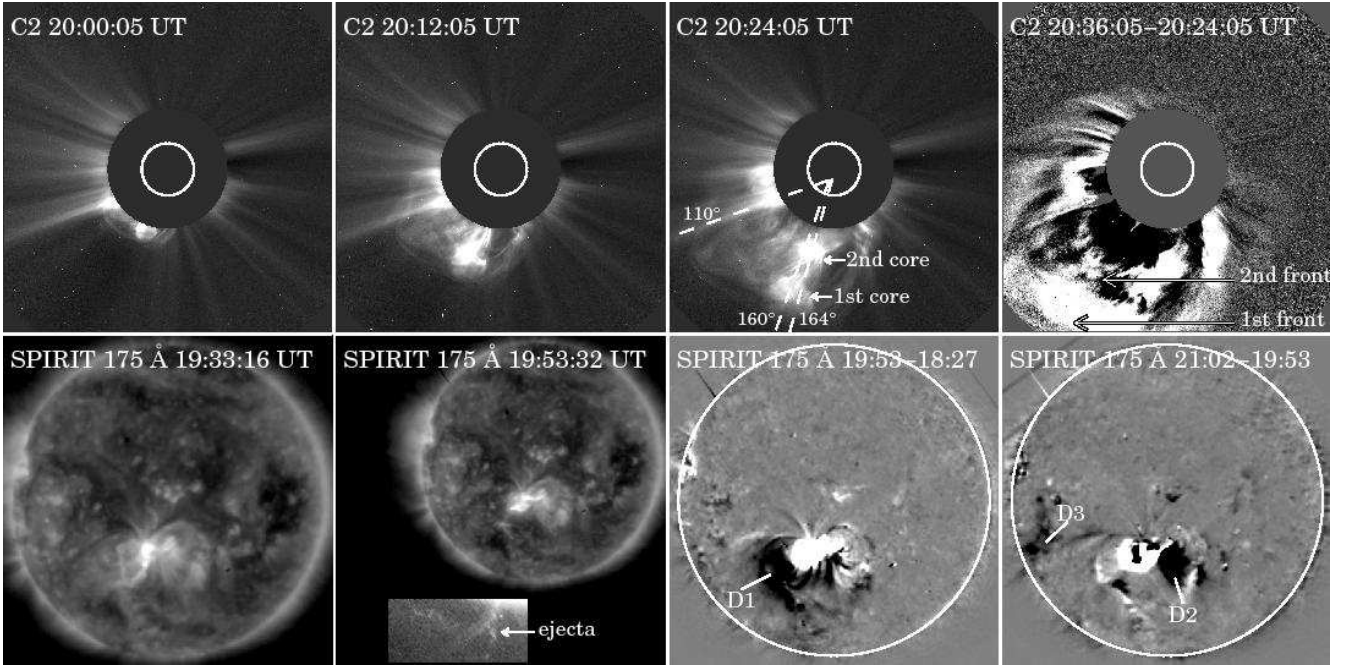}
\caption{{\it Upper panels}: LASCO C2 images showing the evolution of the halo CME. An unusual two-core structure and the associated two bright fronts are conspicuously seen at 20:24~UT and the running difference image at 20:36~UT. The dashed lines overplotted represent the measured position angles of 110\dg\ for the CME flank along a strong streamer structure lying in the southeast, 160\dg\ for the first core (originated from the filament $f2$), and 164\dg\ for the second core (originated from the filament $f3$). {\it Lower panels}: \spirit~175~\AA\ images and their difference showing the ejecta and the coronal dimming regions $D1$ and $D2$/$D3$ after the launch of the first and second CMEs, respectively. \label{fig4}}
\end{figure*}

From the time-lapse movies of both \ha\ center and wing images, it could be seen that the initial flare I was preceded by the eruption of filament $f1$ from the northeastern part of the PIL. Observational evidence for direct cause of eruptions is crucial in understanding their triggering mechanism, while the above fact has been neglected in all the previous studies. Since any upward motion of dark features in \ha\ wavelength is more pronounced in its blue wing, we cut a slit that lies in the projected direction of $f1$ eruption (slit I in the frame 19:12:09~UT), and show the distance-time profile along it in Figure~\ref{slit}$a$. It is unambiguous that $f1$ started to move upward as early as from \sm19:00~UT, with a projected speed of \sm12~\kms. Bright ribbons $1a$ and $1b$ of flare I (see image at 19:23:11~UT) appeared immediately after the disappearance of $f1$ and exhibited the normal expansion motion. The brightenings also readily expanded along the entire PIL mainly toward south. In extreme-UV (EUV) images, the flare brightenings seemingly propagated along the large-scale magnetic field lines (see images from 19:24--19:38~UT). By comparing the preflare EUV image with MDI magnetogram (e.g., Fig.~\ref{fig0}$ac$), these magnetic loops span from the negative magnetic polarity region N1 to the positive magnetic polarity region P1 (cf. Fig.~11 of \citealt{nagashima07}). The distance-time profile (Fig.~\ref{slit}$b$) along the slit II illustrated in Fig.~\ref{fig0}$b$ also shows that this propagating brightening forms immediately after flare I brightenings reach the southeastern portion of the S-shaped PIL. By \sm19:35~UT, the entire field lines were brightened (also see the EUV image at 19:33:16~UT in Fig.~\ref{fig4}) and rapidly erupted outward, together with the co-spatial filament $f2$ at \sm19:38~UT by breaking away from the southern ends. During the flare II, a new pair of separating ribbons $2a$ and $2b$ (denoted in image at 19:51:09~UT) were observed underneath the outward moving $f2$, although this flare can not be obviously recognized in SXR flux. We use the same notation for the flares I and II as in \citet{wang07a}, where the dynamics of flare ribbons and filament $f2$ were studied in details. An important feature here is that Figure~\ref{slit}$b$ clearly demonstrates that $f4$, another filament associated with the southeast portion of the PIL (see Fig.~\ref{fig0}$b$), was not disturbed but remained intact after the flares I and II, which is also evidenced by comparing the pre- and postflare EUV images \citep{wang07a}. Although the slit II only cuts through one segment of $f4$, similar results can be obtained for other parts, which indicate that the whole $f4$ did not erupt. This is in contrast with \citet{nagashima07} where the activity of $f4$ was incorrectly identified. Note that the current position of the slit II was chosen to better reveal the possible linkage between the flares I and II.

In summary, our analysis indicates that filament $f1$ erupted but $f4$ did not, which is important for judging the whole eruption scenario. We will further discuss the significance of these points in \S\S~\ref{onset} and \ref{first}.

\subsection{The Flare III (\sm19:45--20:21~UT)} \label{third}
After the flare I/II, the flaring loops in EUV images are seen to extend westward from the northeast corner of the PIL, and subsequently the flare III was triggered at the \dt\ spot region. This is corroborated by the facts that strong magnetic reconnection in this phase began with the high energy burst $S0$ (see Fig.~\ref{fig1}) where the type IIId burst accelerated in high density (low altitude) with the corresponding microwave sources above the \dt\ spot (see image at 19:51:09~UT), and that HXR emitting sources are present there during the flare III. Although it is barely see in Figure~\ref{fig2}$a$, the elongated ribbons of the flare I evolve to three \ha\ kernels (denoted as $k1$--$k3$). We identify them with flare kernels, because (1) they are compact feature newly brightened during the flare III, and (2) they are co-spatial with HXR/microwave emissions at the times of the high energy bursts $S0$--$S2$ (see images at 19:51:09, 20:00:17 and 20:04:10~UT in Fig.~\ref{fig2}$a$). We note that unlike the ordinary separating ribbons as seen in the flares I and II, kernels $k1$--$k3$ did not show obvious motion during the flare III. Moreover, the co-spatial footpoint-like high energy emitting sources only appeared at times of the bursts $S0$--$S2$. Around the flare peak, the 25--50~keV X-ray emission mainly originated from a source located between the \ha\ kernels $k1$ and $k2$ (see images at 19:56:09 and 20:02:12~UT). This source is most probably a looptop source over the \dt\ spot, since it is spatially aligned with the lower energy X-ray emission in 12--18~keV. We analyzed the corresponding \hsi\ X-ray spectrum of the looptop source using the Object Spectral Executive. The results show that nonthermal contribution dominates in \sm25--50~keV, since the spectrum in this energy range can only be modeled with a nonthermal power-law distribution. Light curves of the \ha\ kernels and the physical nature of the \ha/HXR sources will further be discussed in \S~\ref{kernels}.

We also identified erupting filament and flux loops in \ha\ and EUV wavelengths, respectively, during the flare III. At least part of the filament $f3$, which consists of two strands (see Fig.~\ref{fig0}) that share a common eruptive motion, began to move outward from \sm19:45~UT (see images from 19:51:09--20:02:12~UT) in a similar direction as that of $f2$. As the feature is relatively weak and the projected trajectory does not follow a straight line, the eruption dynamics of $f3$ is best seen in the time-lapse movies. From \sm19:57~UT, a series of moving bright loops ($L1$ at 19:59:14~UT) stemmed from the flaring site at the \dt\ spot. Erupting loops with a larger scale ($L2$ at 20:00:59~UT) were also observed from \sm20:00 UT. The western ends of these loops can be traced back to the flare kernel $k3$ with negative magnetic polarity. Their eastern ends thus should have positive magnetic polarity but were obscured by the bright postflare arcades of the flare II.

\subsection{A High Speed Flow in EUV}
In \trace\ 195~\AA\ images, we find an interesting flow during the rapid rising phase of the flare III SXR emission. To show this flow we made time-sequenced images using a slit area (15\arcsec~$\times$~90\arcsec; illustrated in 19:56:14~UT) around the flow and display them in Figure~\ref{fig2}$b$. Three flow episodes can be identified and their initiations are pointed out with numbered arrows. Among them, the flow 2 is most clearly seen as bright blobs moving at an apparent speed of $\sim$350~\kms. By comparing the timing of the flows with that of the event light curves, it is obvious that the flows seem to be initiated at times when the bursts of nonthermal emissions occurred. The flow can be traced in EUV images until \sm20:04~UT, after which \trace\ observed with a cadence up to $\sim$2 minutes and was affected by ``snow storm'' due to high-energy particles. As a comparison, the bright EUV flow has no obvious counterpart in \ha\ wavelength. However, it is apparent that a small filament (labeled $f2'$) moved in a co-spatial path at the times of the flow (see images from 19:51:09--20:02:12~UT and the movies). Here $f2'$ is a small portion of $f2$ at its western end that was not disturbed during the flare II. After \sm20:02~UT when the blobs of the flow 2 in EUV reached the \dt\ spot region, the northern tip of the flow began to brighten in \ha\ (see image at 20:16:09~UT). We believe that this flow represent streams of enhanced density traveling toward the flaring arcades overarching the \dt\ spot (see images from 19:56:14--20:02:44~UT) explicitly the looptop HXR source.

\subsection{The Flare IV (\sm20:21--20:57~UT)}
During this phase, a new flare kernel $k6$ in \ha\ originated from the brightened northern end of the flow in the region of southern sunspot umbra, and moved westward toward the region of previously brightened kernel $k2$ (see images from 20:16:09--20:44:10~UT and the movies). There simultaneously appeared a conjugate \ha\ ribbon ($k4$), which was activated in its northern part from \sm20~UT (see \S~\ref{kernels}) and expanded later on. Each of the \ha\ kernel/ribbon $k6$ and $k4$ began to have co-spatial HXR emissions from around the time of the burst $S3$ (see image at 20:27:11~UT), and they separated from each other as usual. By \sm20:44~UT, thermal X-ray sources and loop of arcades in EUV overlied the entire flaring region, with HXR sources at the west and east sides. No more flaring activity is seen in the next \sm3 hours.

\section{THE SUCCESSIVE CMES} \label{cme}
We use EUV and coronagraph images and radio dynamic spectra to investigate the morphology and dynamics of the associated CMEs in this event.

\subsection{General Morphology and Coronal Dimmings}
The first image showing the eruptive signature off the limb is captured by \spirit\ in its 175~\AA\ channel at 19:53~UT (see Fig.~\ref{fig4}), showing an ejecta propagating in the southeastern direction. The following development of an asymmetric full-halo CME was first observed in LASCO C2 at 20:00~UT as a bright loop front preceded by a diffuse envelope above the southeast limb. At 20:12~UT, a traditional three-part structure of CMEs can be clearly seen. The C2 occulter was nearly completely surrounded at 20:36~UT by faint extensions in all directions, with the leading edge on southeast being already past the C2 field of view while that on northwest being just above the occulter. Although the LASCO CME catalog \citep{yashiro04} reported this event as a single CME eruption, there is an unusual second core with associated bright front below the first CME core, as conspicuously seen at 20:24 and 20:36~UT. The event was observed in C3 from 20:18 to 23:18~UT, but the two cores can not be resolved, probably because they became diffuse and merged into an extended bright structure at that time. The event was supposed to produce an intense geomagnetic storm since it originated close to the disc center; however, the interplanetary deflection probably made the CME only graze the Earth \citep{wang06b}.

We looked for a coronal dimming, an important signature of CMEs, which usually suggests loss of the coronal mass that are swept into CMEs along opened field lines \citep[e.g.,][]{thompson00}. The difference images between pre- and post-eruption states of the flares I/II and the flare III are presented using \spirit\ observations at 175~\AA\ (Fig.~\ref{fig4}, {\it lower panels}). After the flare I/II, a strong dimming region ($D1$) appears in the southeast of the active region and extends to the west. Within temporal resolution, the fast darkening of D1 at \sm19:48~UT \citep{slemzin06} agrees with the eruption of the brightened EUV loop and the associated filament $f2$ from \sm19:38~UT. Importantly, another strong dimming region $D2$ appears in the west after the flare III, which implies an eruption of a separate CME. Note that although stifled by the bright flare emission, most probably there should also have coronal dimming east of D2, which together represent the consequence of eruption of the large EUV loops $L1$ and $L2$ and the associated $f3$ during the flare III. As a support to this view, there is a dimming region (D3) close to the west limb, which is located in negative magnetic polarity region. Following the picture of \citet{mandrini07}, the expanding western legs of $L1$/$L2$ rooted in positive magnetic fields could reconnect with the fields at D3, and thus produced extending dimming region there.

\subsection{Radio Signatures} \label{radio}
Radio bursts are known to be able to provide valuable clues about the CME eruption as it propagates outward in the solar atmosphere. It is well accepted that type II bursts are the manifestation of shock waves in the middle to high corona usually associated with either large flares or fast CMEs \citep{nelson85}, and the type II precursors \citep{klassen99} are a signature for the onset of shock formation in the low corona due to expanding of CME structure \citep[e.g.,][]{dauphin06}. According to Figure~\ref{fig1}, radio bursts in this event started at 19:23~UT with broadband pulses in the \sm0.7--18~GHz range and lasted till 19:35~UT. At lower frequencies (\sm150--400~MHz), this initial impulsive phase is followed by drifting radio features during 19:38--19:45~UT. These can be considered as a precursor of a weak but clearly identified type II radio burst, which began with starting frequency \sm40~MHz and drifted toward \sm20~MHz at 19:48~UT (marked with dotted line).

We also note a drifting pulsation structure \citep[DPS;][]{kliem00} during 19:37--19:42~UT in the frequency range of 600--800~MHz. This negatively drifting DPS1 (\sm$-0.7$~MHz~s$^{-1}$) indicates plasmoids formed below the upward rising flux rope. \citet{karlicky07} claimed that plasmoid forms in the phase of intense electron acceleration. This is supported by the simultaneous observation of DPS1 and a strong type III radio burst, which is conventionally interpreted as accelerated electrons escaping along open field lines.

It is worth mentioning that the type IIId burst and the followed normal type III burst is observed at the start of a positively drifting DPS2 (\sm$4$~MHz~s$^{-1}$) during 19:51--19:55~UT in the 1--2~GHz range, which suggests plasmoids propagating downwards in the solar atmosphere. It can also be seen that the radio emission in the \sm600--1300~MHz drifts as a whole toward higher frequency (DPS3) during 20:13--20:22~UT, when the flare IV began. We will see in the following sections that these emission features in the radio dynamic spectra are consistent with our interpretation of the successive eruptions.

\begin{figure}
\epsscale{1.15}
\plotone{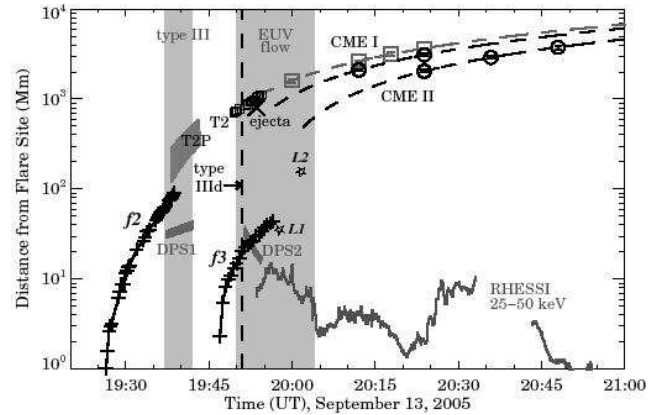}
\caption{Time evolutions of the distances of the erupting filaments $f2$ and $f3$, type II radio burst (T2) and its precursor (T2P), DPSs, and CMEs. The three dashed lines from up to down are least-squares linear fit to the data points of the flank of the CME I, the center of the first core, and the center of the second core (CME II). The derived velocities for the CME I flank, the bright cores of the CME I and II, and the type II radio burst are \sm1430, 1390, 1190, and 1390~\kms, respectively. The approximate positions of the erupting EUV loops $L1$ and $L2$ are denoted. The timings of other activities (type III and IIId bursts, and EUV ejecta and flow) are also illustrated. \label{fig5}}
\end{figure}

\subsection{Eruption Kinematics}
In order to link the CME activities with their source regions on the solar surface, height-time measurements of erupting associated features are carried out using the following scheme. We approximate the height from the projected distance of the ascending filaments ($f2$ in \trace\ 195~\AA\ and $f3$ in BBSO \ha~$-$~0.6~\AA), the flux loops ($L1$ and $L2$), and the EUV ejecta, and compare them with those of the CME features in the plane of the sky, which include the two bright cores (centroid positions are measured) and the CME flank (where it interacted with a streamer at southeast). Due to the difficulty to follow erupting loops near bright flaring region and the limited field of view of \trace, the averaged positions of $L1$ and $L2$ are used. In addition, the heights of the metric type II radio burst and the type II precursor are estimated using Mann's and fourfold Newkirk's coronal electron density models, respectively, following \citet{warmuth05}, and those of the DPSs at lower corona are approximated using the model of \citet{aschwanden02}.

Based on these results presented in Figure~\ref{fig5}, we organize the observational evidence for magnetic eruption as follows. First, it is obvious that the filament $f2$, the type II radio burst and its precursor, the ejecta, and the CME structure first seen in LASCO images (CME I) are closely related. This suggests that (1) the type II precursor signifies the shock formation in the low corona caused by the expanding flux rope, (2) the type II radio burst formed at the CME flank when interacting with the dense streamer (i.e., low Alfv\'enic region), as previously reported \citep[e.g.,][]{cho07,liu07b}, and (3) the filament $f2$ evolves to become the first core seen in LASCO images. It is also interesting to note that the DPS1 and the associated type III radio burst occurred during the rapid rising phase of the CME, suggesting the ejection of plasmoids and abrupt particle acceleration \citep{karlicky04,karlicky07}.

Second, the origin of the second bright core seen in LASCO images can be traced back to the erupting filament $f3$. It is thus most likely that the filament $f3$ and the erupting bright loops $L1$ and $L2$ produced a separate CME event (CME II), which is closely associated with the flare III at the \dt\ spot region. Another strong evidence is that this second CME is related to the co-spatial strong coronal dimming region $D2$ after the flare III. Note that (1) the heights of $f3$ in its later eruption phase with higher speed could be underestimated since they are measured using \ha\ wing images, and (2) there could have a rapid acceleration of flux ropes around 19:55--20:00~UT when HXR peaked \citep[see e.g.,][]{qiu04b}. It is also shown that the type IIId burst and DPS2 occurred at the beginning of this eruption, and the EUV flow is co-temporal with the rising phase of the CME II.

Third, the two filaments $f2$ and $f3$ initially erupted in a similar direction (\sm146\dg; cf. Fig.~\ref{fig2}$a$ and the movies), and the first and second bright cores of the CME are also aligned radially (\sm160\dg; see Fig.~\ref{fig4}). This corroborates our speculation that the first and second CME cores would be associated with the dense filament materials $f2$ and $f3$, respectively. The deviation in direction, however, is not readily explainable because there lacks of three-dimensional observations of the CME evolution in the low corona.

\begin{figure}
\epsscale{1.15}
\plotone{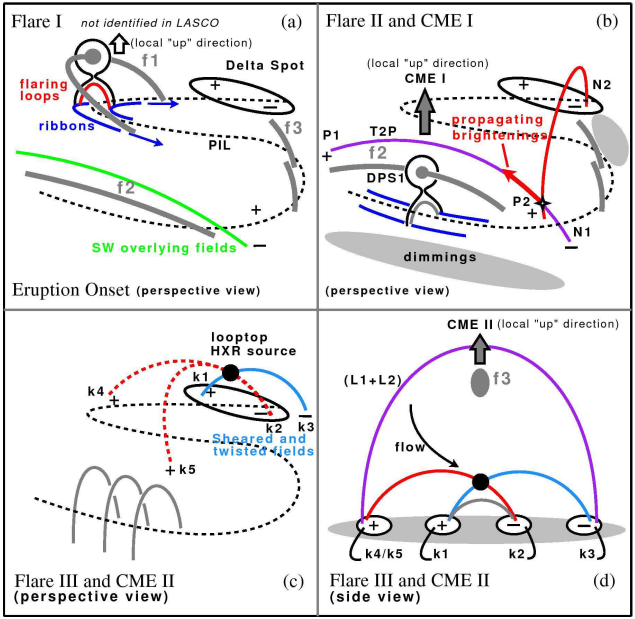}
\caption{Schematic sketches of the main episodes of the eruption based on our observational evidence. The filaments $f1$--$f3$ and magnetic spots P1, P2, N1, and N2 in {\it a} and {\it b} are those denoted in Fig.~\ref{fig0}. The flare kernels $k1$-$k5$ and erupting loops $L1$/$L2$ in {\it c} and {\it d} are corresponding to those in Fig.~\ref{fig2}$a$. Note that signature in LASCO images for $f1$ eruption can hardly be identified. See detailed discussions in \S~\ref{idea}. \label{fig55}}
\end{figure}

\section{SPECULATION ON SUCCESSIVE FLARES AND CMES} \label{idea}
We here argue that these series of flares and CMEs are interrelated to each other rather than separate individual events due to a simple coincidence of occurrences, using the schematic illustration shown in Figure~\ref{fig55}.

\subsection{Eruption Onset}\label{onset}
A careful examination of \ha\ images reveal that the event began with the eruption of the filament $f1$ from the northeastern part of the PIL (Fig.~\ref{fig55}$a$). This was not reported in the previous studies. On one hand, the C- and M-class flares that occurred in the nearby region during one day before the X1.5 event \citep{nagashima07} could have possibly led to the loss of equilibrium of $f1$. On the other hand, \citet{chifor07} found preflare traveling HXR sources during \sm19:10--19:20~UT along the PIL in the vicinity of $f1$, which appear co-temporal with the rising of $f1$. Thus it is also possible that the $f1$ eruption is driven by the tether-cutting reconnection manifested as these HXR sources. Our data, at this stage, cannot unambiguously distinguish the above two possibilities \citep[cf.][]{moore06,liu07b}. As the filament moving upward, it could sequentially tear away the overarching magnetic fields \citep{tripathi06b} in a larger scale, and accordingly we see that the flare ribbons expanded bi-directionally towards the entire PIL. We found it hard to detect coronal dimming, if any, associated with the eruption of $f1$ because the nearby flare brightening was strong and the cadence of the full-disk coronal images was low. Neither could we find the corresponding signature of $f1$ in the coronagraph images, probably because $f1$ erupted in a different direction to $f2$ and $f3$ (cf. orientation of the slits I and II), which could make it unfavorable to be detected.

We note, however, that in \citet{chifor07} the eruption of $f2$ and the high energy emissions at \sm20~UT were considered as direct consequence of the precursor seen as the preflare traveling HXR sources. Based on detailed observational evidence, we rather suggest that $f1$ and the subsequent flare I are directly related to the HXR precursor, because (1) $f1$ is much more closer than $f2$ to the preflare HXR sources, and (2) the flare emissions at \sm20~UT are clearly a completely separate energy release episode occurred about 40 minutes later.

\subsection{The Flares I, II, and CME I}\label{first}
The most interesting feature during the flare I is the propagating brightening along the large-scale magnetic field lines. \citet{chifor07} interpreted this brightened loop as the eruption of the filament $f2$; however, $f2$ is clearly seen as a separate darker structure that rose together with the bright loops (also see larger images in Fig.~3 of \citealt{wang07a}). Alternatively, \citet{nagashima07} attributed the bright loop structure to the eruption of another filament $f4$, which lies under $f2$ along the southeast PIL. Using the OSPAN images at \ha\ center wavelengths, we have shown clearly that only $f2$ erupted and $f4$ remained undisturbed. The same conclusion was reached by \citet{wang07a} based on EUV images. \citet{wang07a} speculated that the traveling brightening is due to thermal conduction or ejected hot chromospheric plasma resulted from heating by the extended flare ribbon emissions. However, in many other events of ribbon expansion, such a phenomenon is not usually observed.

We here propose a scenario that could account for this rare phenomenon based on our observations. It is well known that in the standard flare model, a flare ribbon in one magnetic polarity is paired with a ribbon in the other magnetic polarity region, and both connect to the coronal X-point. During the flare I, the ribbon that extended southward to become a curved J-shape lies in the positive magnetic polarity region. This ribbon must pair with the ribbon in negative magnetic field around the southern umbra. Hence, there should have been flaring magnetic fields connecting from the region P2 to N2 (denoted in Fig.~\ref{fig0}) and to further western region ({\it red} in Fig.~\ref{fig55}$b$), which are evidenced by the elongated microwave source seen at 19:24:11~UT in Fig.~\ref{fig2}$a$ and are also discernible in EUV images. Since the flare region was quickly expanding, the fields P2-N2 could be forced to reconnect with the close-by large-scale fields N1-P1 at \sm19:24~UT. The reconnection could be low in the chromospheric level, and the materials heated mainly by the flare I can subsequently rise into and propagate along the cold magnetic fields N1-P1, as described in \S~\ref{initial}. Idea of such a forced reconnection between expanding flaring magnetic fields with other favorably oriented magnetic structures was put forward by several authors \citep{goff07,demoulin07,van08}.

An immediate consequence of the above reconnection is the decrease of magnetic tension that presses down the filament $f2$ lying below or being part of the flux loops N1-P1. Since the equilibrium state of $f2$ was already changed during its slow ascending motion over two days before the event \citep{nagashima07}, it began to erupt outward and created two separating ribbons underneath (flare II) in the standard way (Fig.~\ref{fig55}$b$). Later on, the filament $f2$ might completely lose its equilibrium \citep{forbes95} around 19:38~UT, and pushed open the overlying fields (including P1-N2 resulted from the reconnection) to become the CME I. The extending coronal dimming region from P1 to N2 and further west strengthens the idea of involvement of these regions via the reconnection scenario proposed above. On the contrary, \citet{nagashima07} claimed that the catastrophic eruption of $f2$ was triggered by a small C2.9 flare at 19:05~UT. Our interpretation is similar to the picture of \citet{wang07a}, in which the destabilization of $f2$ was ultimately due to the initial flare originated from the northeast corner of the PIL. The C2.9 flare contributed to the loss of equilibrium of $f2$, but might not be the direct cause.

\subsection{The Flare III and CME II} \label{kernels}
The results of imaging and timing analysis clearly demonstrate that the CME II is closely associated with the flare III at the \dt\ spot region with strong magnetic fields. It occurred after the eruption of $f2$, which lies around the southeastern part of the PIL in a weaker fields. Since $f2$ seems not to be rooted at the \dt\ spot region, it is unlikely that the filament eruption is directly related to the flare emissions at the strong fields, which instead is the case in \citet{sterling04}. We note that the magnetic fields at the \dt\ spot region might be highly sheared and twisted, as the main positive and negative polarities had a fast counterclockwise rotation since they appeared from the west limb \citep{nagashima07,li07}. These sheared core fields could erupt outward after the overlying magnetic fields being removed by the CME I as evidenced by the extending coronal dimming, a scenario similar to the breakout model \citep{antiochos99} and which has also been observed in some other events \citep[e.g.,][]{gary04,sterling04}.

\begin{figure}
\epsscale{1.15}
\plotone{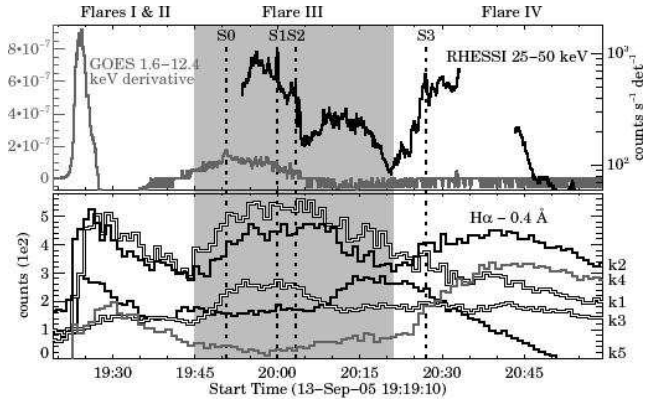}
\caption{Light curves of average \ha~$-$~0.4~\AA\ emission for the flare kernels $k1$--$k5$ determined in areas marked in Fig.~\ref{fig2}$a$, in comparison with the time profiles of \hsi\ HXR emissions and the time derivative of \goes\ SXRs. The bursts $S0$--$S3$ are those denoted in Fig.~\ref{fig1}. For display purpose the values of $k3$, $k4$, and $k5$ was subtracted by 100, 120, and 320, respectively, and those of $k2$ was added 60. \label{fig6}}
\end{figure}

To study the eruption of the sheared core in more detail, we plot in Figure~\ref{fig6} the light curves of the three most obvious flare kernels ($k1$--$k3$ in Fig.~\ref{fig2}$a$) in \ha~$-$~0.4~\AA\ and compare them with that of HXRs. The results show that $k1$--$k3$ have similar emitting time profiles as HXRs during the flare III. In particular, co-temporal peaks in \ha~$-$~0.4~\AA\ are discernible when the bursts $S0$--$S2$ occurred. As blue wing \ha\ images best represents the precipitation of energetic HXR-emitting electrons in the lower atmosphere \citep[e.g.,][]{lee06}, this further demonstrates that $k1$--$k3$ are the footpoints of the flare III. Interestingly, co-spatial HXR footpoint sources only appear at the times of the bursts $S0$--$S2$. We here note that from \hsi\ observation, nonthermal looptop sources can occur when column density in the coronal loop is sufficiently high \citep{veronig04}. In this event, EUV images show very strong emissions from top of the low-lying interacting loops above the \dt\ spot and there is a co-spatial and persistent looptop HXR source, both implying high densities there. Detailed examination of the loop density is however out of the scope of this study.

Although some flares show three footpoints \citep[e.g.,][]{hanaoka99}, they are often non-eruptive. In the present event, co-temporal erupting EUV loops $L1$/$L2$ clearly originated from the source region of the flare III, which usually suggests a quadrupolar magnetic configuration \cite[e.g.,][]{yurchyshyn06b}. As the fourth footpoint should be in the positive magnetic field region, we try to identify it by overplotting in Figure~\ref{fig6} the light curves of the only other two flare kernels $k4$ and $k5$ with positive magnetic polarity (see image at 20:16:09~UT in Fig.~\ref{fig2}$a$). The results show that the time profile of $k5$ may have a stronger correlation with those of $k1$--$k3$, while $k4$ also began to brighten from around the HXR peak. The two kernels are also similar in that there are hardly emissions in HXRs, although co-spatial HXR sources with weak intensity ($<$30\% of the peak intensity of the co-temporal strong sources near the \dt\ spot) can be detected for $k4$ and $k5$ while only at \sm20:01 and 20:04~UT, respectively. Considering both $k4$ and $k5$ are spatially feasible to be the eastern foot of the loops L1 and L2 (see images at 19:59:14 and 20:00:59~UT in Fig.~\ref{fig2}$a$), it appears that both of them could be associated with this flare. As to why there lack of HXR/\ha\ emissions, we offer the following speculation. As this flare is heavily involved with type III radio bursts, it suggests a close involvement of open field lines. According to the results of potential field source surface model\footnote{See, e.g., \url{http://www.lmsal.com/forecast/TRACEview/images/TRACEfov_20050913_185959.tiff}}, open field lines of this active region lie east to the \dt\ spot concentrating in the region spanning from $k4$ to $k5$. Thus accelerated electrons, instead of precipitating downward, could drift to and hence escape from the ambient open field lines \cite[cf.][]{liu06}.

Therefore, we envision a picture, in which field lines $k4$/$k5$--$k2$, representing the expanding flaring fields from the flare I, reconnect with the sheared fields $k1$-$k3$ at the \dt\ spot region. The looptop HXR source is associated with the site of the reconnection (Fig.~\ref{fig55}{\it c}). The outcome of this reconnection is the observed erupting loops $L1$ and $L2$, which could also be pushed outward by the filament $f3$ and together formed the CME II (Fig.~\ref{fig55}{\it d}). It is also consistent that the footpoints in this flare are nearly fixed, in contrast to normal separating ones as a result of reclosing of field lines.

\subsection{Nature of the High Speed EUV Flow}
In the above context of magnetic reconnection, it is meaningful to investigate the nature of the bright EUV flow associated with the rapid rising phase of the CME II based on our observations. First, the flow was brightened up at its northern tip in \ha\ when it reached the \dt\ spot region. If this indicates heating possibly due to flow-driven compression, the flow should move downward in the solar atmosphere. Moreover, the simultaneous positively drifting DPS2 could be a radio signature of the plasmoids moving downward. Second, the flow is co-temporal with nonthermal emissions and travels toward the looptop (Fig.~\ref{fig55}{\it d}) at a high speed. This suggests that the flow could be associated with the magnetic reconnection process of the flare III/CME II. In specific, we identify the observed flow with an inflow toward the X-point as the CME moves away from the reconnection region as in \citet{temmer08} and \citet{aschwanden08b}. Of course, we are not sure whether or not this is a downward flow, because this region is close to disk center, and the signal for the narrow flow in the dopplergrams constructed using \ha\ wing images are weak. But it is worthwhile to compare this observation with others. To our best knowledge, the only other reported bright EUV flow during solar eruption was observed by \citet{tripathi06c}, in which an EUV downflow near the limb is seen in the course of a prominence eruption associated CME. Different from our event, the EUV flow coincided with the deceleration phase of the CME, and was thus considered as materials sliding down along contracting magnetic arcades formed by the reconnection.

However, we note that this EUV flow can not be regarded as coronal rains because the observed apparent speed is higher than that of ordinary coronal rain \cite[50--100~\kms;][]{tandberg77}. Neither can it be gravitationally falling material from the erupting filament $f2$, because materials would have to accelerate over a distance of 300\arcsec\ starting from rest at about 19:38~UT, if we take 350~\kms\ as the final speed under constant gravitational acceleration. Although it exceeds the \trace\ field of view, no signatures of falling materials from high altitude can be visualized up to the time of the flow. We would like to suggest that (1) the reconnection inflow is the most likely explanation for the EUV flow, and (2) the heated materials of close-by preceding flarings might be the reason that the inflow is visible in this particular event.

\subsection{The Flare IV}
After the lift-off of CME II, the flaring site moved east from the \dt\ spot region. The flare ribbons $k4$ and $k6$ separated from each other, with $k6$ moving from the northern end of the flow into the former $k2$ region. For the time being, we can only suspect that the CME II might have teared away nearby magnetic fields in its way out \citep[cf.][]{balasubramaniam05}, and the newly brightened separating kernels could just represent the reclosing of such opened field lines. Meanwhile, the positively drifting (downward moving) feature DPS3 in radio dynamic spectrum at the beginning of this flare also remains a puzzling issue. As related studies, \citet{barta08} showed that the plasmoid formed after reconnection can move upwards as well as downwards in dependence on the surrounding magnetic fields. For the downward moving plasmoid, it can further interact with arcades below and the additional energy can be released. Observational evidence of such interaction have been presented by \citet{kolomanski07}.

\begin{deluxetable}{lll}
\tablewidth{0pt}
\tablecaption{TIME LINE OF EVENTS\label{time}}
\tablehead{\colhead{Start Time} & \colhead{Eruption} & \colhead{} \\
\colhead{(minutes)} & \colhead{Episode} & \colhead{Characteristic Features}}
\startdata
$-23$.......... & Onset & $f1$ starts to move upward, HXR precursor\\
$-4$............& Flare I & Flare starts in \goes\ SXRs\\
$0$.............. &       & Expanding flare ribbons along PIL\\
$+1$............ &      & Propagating EUV brightening along field lines\\
$+15$.......... &  Flare II     & Catastrophic eruption of $f2$, DPS1,\\
             &       & type II bursts, dimming, and the first CME\\
$+22$.......... & Flare III  & Stationary flare kernels, looptop HXR source,\\
             &       & EUV flow, DPS2, $f3$ eruption, dimming,\\
	     &	     & and the second CME\\
$+58$.......... & Flare IV  & Separating flare kernels/ribbons and DPS3
\enddata
\tablecomments{Approximate start time is given in minutes from 19:23~UT.}
\end{deluxetable}

\section{SUMMARY AND DISCUSSION} \label{summary}
In this paper, we have presented a comprehensive study of the 2005
September 13 eruption that comprises four flares and two fast CMEs.
We found that these eruptions originated from various locations
along the elongated S-shaped PIL of NOAA AR 10808. Since they occur in one active region within about 1.5 hours, we identified this event with successive flares/CMEs. The event timing and characteristic of each eruption is described in Table~\ref{time}. We summarize
our major findings and interpretations as follows.

\begin{enumerate}

\item The whole event started by the eruption of a filament in the
northeastern  part of the active region. This eruption produced the flare I loops that expanded across the active region along
the PIL. These expanding flare loops interacted with the large-scale
magnetic fields in the southeast region, and heated the materials to be ejected into the cold flux loops as seen in EUV.

\item The underlying filament subsequently expanded outward as the
overlying field weakens. This filament system erupted to become the CME I
and the flare II occurred underneath. The type II precursor represents the shock
formation in the low corona due to expanding flaring loops. 
The negatively drifting DPS1 and type radio II burst represent ejection
of plasmoids and interaction of the CME flank with the dense
coronal streamer, respectively.

\item The CME I could have partially removed overlying magnetic
fields in the northwestern \dt\ spot region as implied by the extending
coronal dimming. As a result, the sheared core fields erupted outward to interact with the flaring loops
of the flare I, which is manifested by the sustained looptop HXR
source and footpoint-like HXR/microwave sources during the three high-energy
bursts of the flare III. Subsequently, the reconnected loops and
another filament erupt to become the CME II.

\item The flare IV shows standard ribbon motion, which indicates
reclosing of magnetic fields opened by the CME II. In addition, we found a fast EUV flow and suggest that it should represent the reconnection inflow toward the reconnection site as the CME II moved outward.

\end{enumerate}

We conclude that the event was initiated by a small disturbance from
a relatively weak magnetic field, in contrast with other eruptions
that initiate from strong magnetic fields such as \dt\ spot
\citep[e.g.,][]{liu05}. This kind of activities evidences a chain
reaction of consecutive activities occurred in a single active
region, which is similar to the so-called domino effect proposed by
\citet{zuccarello09}. The successive flares and CMEs can be
distinguished from the well-known sympathetic flares/CMEs due to
multiple eruptions occurring in different active regions. We believe
that successive flares and CMEs is another challenge for the space
weather research, as it implies restructuring of the coronal
magnetic fields in a much more complex way.

\acknowledgments
The authors thank the teams of GBSRBS, OSPAN, OVSA, \hsi, \soho, \spirit, \trace, and \wind\ for efforts in obtaining the data. We are grateful to the referee for many valuable comments that greatly improved the paper. C.~L. thanks Drs. R. Moore and A. Sterling and other colleagues for constructive discussions, Drs. K. Cho and S. Tun for help with OVSA data, Dr. V. Slemzin for providing the \spirit\ data, and Dr. S. White for providing the GBSRBS data. C.~L. and H.~W. were supported by NSF grants ATM 08-19662, ATM 07-45744, and ATM 05-48952, and NASA grants NNX 08AQ90G and NNX 08AJ23G. M.~K. was supported by grant IAA300030701 of the Grant Agency of the Academy of Sciences of the Czech Republic. D.~P.~C. and N.~D. were supported by NSF grant ATM 05-48260 and NASA grant NNX 08AQ32G. OSPAN is a PI driven project by Air Force Research Laboratory Space Vehicles Directorate (RVBXS) and the National Solar Observatory. The OVSA was supported by NSF grant AST-0607544 and NASA grant NNG06GJ40G to the New Jersey Institute of Technology. \hsi\ and \trace\ are NASA Small Explorers. \soho\ is a project of international cooperation between ESA and NASA. The \spirit\ experiment was carried out by the Laboratory of X-ray Astronomy of the Sun of P.~N. Lebedev Physical Institute, Moscow, Russia, under the \textit{CORONAS} solar investigation project.

\end{document}